\documentclass[11pt]{article}
\pdfoutput=1

\usepackage{jheppub}
\usepackage{bbm}
\usepackage{color}
\usepackage{url,comment}
\usepackage{times}
\usepackage{latexsym}
\def\beq{\begin{equation}}
\def\eeq{\end{equation}}
\def\bey{\begin{eqnarray}}
\def\eey{\end{eqnarray}}

\def\lsim{\mathrel{\raise.3ex\hbox{$<$\kern-.75em\lower1ex\hbox{$\sim$}}}}
\def\gsim{\mathrel{\raise.3ex\hbox{$>$\kern-.75em\lower1ex\hbox{$\sim$}}}}

\newcommand{\be}{\begin{equation}}
\newcommand{\ee}{\end{equation}}

\newcommand{\gev}{{\rm ~GeV }}

\newcommand{\ttbar}{t\bar{t}}
\newcommand{\met}{E\!\!\!/_T}

\preprint{RU-NHETC-2013-17}

\title{Multi-Lepton Signals of Top-Higgs Associated Production}

\date{\today}

\author[a,b]{Nathaniel Craig,} \author[b]{Michael Park,}
\author[c]{and Jessie Shelton} \affiliation[a]{School of Natural
  Sciences \\ Institute for Advanced Study, Princeton, NJ 08540}
\affiliation[b]{Department of Physics \\Rutgers University,
  Piscataway, NJ 08854} \affiliation[c]{Center for the Fundamental
  Laws of Nature,\\ Harvard University, Cambridge, MA
  02138} 
\emailAdd{ncraig@ias.edu}
\emailAdd{q1park@physics.rutgers.edu}
\emailAdd{jshelton@physics.harvard.edu}

\abstract{We evaluate the potential to measure $t \bar t H$ associated
  production at the LHC using non-resonant multi-lepton final states
  in conjunction with two or more $b$-tags. The multi-lepton $t \bar t
  H$ signal arises predominantly from $H \to \tau^+ \tau^-$ and $H \to
  WW^*$ alongside the semi-leptonic or fully leptonic decay of the $t
  \bar t$ pair. We demonstrate the power of a multi-lepton search for
  $t \bar t H$ associated production by recasting the CMS $b$-tagged
  multi-lepton search with 19.5 fb$^{-1}$ of 8 TeV data to obtain an
  observed (expected) limit of 4.7 (6.6) times the Standard Model
  rate, comparable to ongoing searches in $4b$ and $bb\gamma \gamma$
  final states. Sensitivity can be further improved by the addition of
  exclusive channels involving same-sign dileptons.  We recast the CMS
  $b$-tagged same-sign dilepton search with 10.5 fb$^{-1}$ of 8 TeV
  data to set limits on $\ttbar H$ associated production, and
  approximately combine the two searches by calculating the fraction
  of same-sign dilepton signal events which do not satisfy
  multi-lepton selection criteria.  We estimate an expected total
  non-resonant leptonic reach of $\mu < 5.0$ times the Standard Model
  rate in 20 fb$^{-1}$ of 8 TeV data, with improvements possible.  }
 \arxivnumber{NNNN.NNNN}
\begin{document} 

\maketitle

\section{Introduction}

Understanding the properties of the newly-discovered Higgs boson is
one of the major goals of the Large Hadron Collider (LHC) physics
program. One key property of the Higgs is its Yukawa coupling to the
top quark, $y_t$.  While the loop-induced production of the Higgs
through gluon fusion is sensitive to the top Yukawa, as is the Higgs
decay to photons, these processes are subject to potential
interference from particles beyond the Standard Model (SM) running in
the loop.  Only a direct measurement of associated $t\bar t H$
production can unambiguously be used to determine the top Yukawa
coupling.

The favored channel for observing $t\bar t H$ at the LHC exploits the
dominant Higgs decay to $b\bar b$ \cite{ATLAS-CONF-2012-135,
  Chatrchyan:2013yea}. While this channel benefits from the large
$h\to b\bar b$ branching fraction, the combinatoric difficulty of
reconstructing the $4b$ final state means measurements of $t\bar t H$
are not straightforward.  Despite sophisticated multivariate analysis
techniques and new ideas for handling combinatorics
\cite{Plehn:2009rk, Artoisenet:2013vfa}, the $b$-quark decay of the
Higgs remains a challenging signal.  With 5.0 fb$ ^ {-1}$ of data at 7
TeV and 5.1 fb$ ^ {-1}$ at 8 TeV, the CMS experiment excludes a 125
GeV SM Higgs at $\mu = 5.8$ times the SM expectation in the $4b$ final
state \cite{Chatrchyan:2013yea}. More recently, by combining the $H
\to b\bar b$ and $H \to \tau^+ \tau^-$ final states with 19.5
fb$^{-1}$ of 8 TeV data, the CMS experiment excludes a 125 GeV SM
Higgs in the $t \bar t H$ channel at $\mu = 5.2$ times the SM
expectation \cite{CMS-PAS-HIG-13-019}. Meanwhile, the clean but rare
decay $H\to \gamma\gamma$ constrains the process $pp \to t\bar t H$,
$H\to\gamma\gamma$ at $\mu = 5.4$ times the SM expectation using 19.6
fb$^{-1}$ at 8 TeV at CMS \cite{CMS-PAS-HIG-13-015} and at $\mu =5.3 $
times the SM in 20.3 fb$^{-1}$ at ATLAS \cite{ATLAS-CONF-2013-080}.

The subdominant decay modes of the Higgs into the leptophilic
$WW^{*}$ \cite{Maltoni:2002jr, Kostyukhin:685390} and $\tau^+
\tau^-$ \cite{Belyaev:2002ua} final states offer a different path
toward the observation of $t\bar t H$ associated production.
Including leptons from the decays of the top quarks, $t\bar t H$
events followed by $h\to WW^*$ and $h\to \tau^+ \tau^-$ yield up
to four charged leptons\footnote{The small Higgs branching fraction into
  partially and fully leptonic $ZZ^*$ contributes negligibly to
  multilepton reach, but can in principle give final states with as
  many as six charged leptons.}.  Since the charged leptons (including
hadronic $\tau$s) are accompanied by an equal number of neutrinos,
full event reconstruction is generally not straightforward. However,
as multi-leptonic final states are relatively low-background and
well-understood, {\em non-resonant} multi-lepton searches offer a
powerful handle on $t\bar t H$ associated production.

LHC searches for multi-leptonic final states obtain excellent reach
for new physics from combining many exclusive channels
\cite{Chatrchyan:2012mea, CMS-PAS-SUS-12-026, CMS-PAS-SUS-12-027,
Chatrchyan:2013xsw, ATLAS:2013rla, ATLAS:2013qla,
ATLAS-CONF-2013-070}. The sensitivity of these non-resonant
multi-lepton searches to Higgs physics both in and beyond the SM has
already been demonstrated \cite{ContrerasCampana:2011aa, Craig:2012vj,
Craig:2012pu, Chang:2012qu}.  Here we show that, with the addition of
$b$-tagging information, nonresonant multi-lepton searches can also
begin to constrain the challenging $t\bar t H$ associated production
process, at a sensitivity nearly equal to that of $H\to b\bar b$ and
$H\to \gamma\gamma$.  We consider here $3\ell$ and $4\ell$ final
states, arising from {\em both} $h\to\tau^+ \tau^-$ and $h\to WW^*$,
assuming the SM value for the ratio of partial widths.  After
accounting for additional background contributions, searches from
multi-leptons alone limit associated $\ttbar H$ production at $ \mu <
4.7$ times the SM rate at $95 \%$ CL.  While this limit benefits from
low background fluctuations, adding exclusive same-sign dilepton
channels \cite{Maltoni:2002jr, Curtin:2013zua} could significantly enhance
the reach. 

Observation of $t\bar t H$ in multi-leptons allows an additional novel
measurement of $y_t$ which is both independent of $y_b$ and subject to
different systematics.  Thus the search proposed here provides an
important new handle on the coupling of the Higgs to fermions,
rounding out the picture of electroweak symmetry breaking that is
emerging from LHC data.

The organization of this paper is as follows. In
Section~\ref{sec:multilep} we use CMS multi-lepton searches to set
limits on SM $\ttbar H$ associated production.  In
Section~\ref{sec:ssdilep} we discuss the additional reach available in
same-sign dilepton searches, and in Section~\ref{sec:conclusions} we
conclude.

\section{Signatures of $\ttbar H$ in multileptons}
\label {sec:multilep}

\subsection{Multi-lepton search strategy}

The multi-lepton search strategy pursued by CMS to great success
partitions events with three and four identified leptons into multiple
exclusive bins, using all bins (including those with low
signal-to-background) to set limits on physics beyond the SM. This
procedure is flexible and offers sensitivity to many models of new
physics.

The major SM multi-lepton backgrounds come from electroweak diboson
production.  Processes with a fake isolated lepton, mainly $\ttbar$
and inclusive $Z^{(*)}$ production, also contribute significantly.
Asking that events have one or more $b$-tag dramatically reduces the
dominant SM multi-lepton background $WZ$, as well as $Z^{(*)}+X$. In
events with one or more $b$-tag, the dominant backgrounds are reduced
to $t\bar t$, except in bins with four light leptons, where $ZZ$ and
$t\bar t Z$ dominate. The recent multi-lepton searches in $b$-tagged
events are sensitive to fb-level cross sections
\cite{Chatrchyan:2013xsw}.

Here we recast the $b$-tagged multi-lepton search of
\cite{Chatrchyan:2013xsw} as a limit on $t\bar t H$ production.  This
search, performed with 19.5 fb$^{-1}$ of data at 8 TeV, divides three-
and four-lepton events into several exclusive categories, based on the
following variables:
\begin {itemize}

\item Presence or absence of a hadronic $\tau$.  Bins with exactly 0
  or 1 hadronic $\tau$ are considered.

\item Number of leptons (including hadronic $\tau$s).  The included
  bins are those with exactly 3 or exactly 4 leptons.

\item Event $S_T$, defined as
\beq
 S_T = \sum_{\mathrm {leptons}}|p_T| +\sum_{\mathrm {jets}} |p_T| + \met .
\eeq
Events are divided into five bins with $S_T$ ranging from 0-300 GeV,
300-600 GeV, 600-1000 GeV, 1000-1500 GeV, and $> 1500$ GeV.
\end {itemize}
All events in this analysis are required to have at least one
$b$-tagged jet.  Any opposite sign, same flavor (OSSF) pair of leptons
must also satisfy $m_{\ell\ell}>12$ GeV to suppress contributions from
$\gamma^*$ and hadronic bound states, and $|m_{\ell\ell}-m_Z|>15$ GeV
to suppress backgrounds with a $Z$.

\subsection{Event generation and simulation}
\label{sec:evtgen}

Associated $t\bar t H$ events are simulated in MadGraph 5
\cite{Alwall:2011uj}, showered in Pythia \cite{Sjostrand:2006za}, and
run through the PGS detector simulator \cite{PGS}.  Separate samples
of 40000 events are generated independently for each lepton
multiplicity in each included Higgs decay channel to ensure good
statistical coverage.  Tops and Higgs bosons are decayed in MadGraph.

Our implementation of PGS has been tuned to better approximate the
performance of the CMS detector.  In particular, muon and hadronic tau
isolation routines have been modified in order to obtain better
agreement with CMS results.  Separate variables parameterizing the
isolation of muon and tau candidates respectively are defined as
follows. Defining $\Delta R_{i, \mu}$ to be the distance in $\eta -
\phi$ space between an object $i$ and a muon candidate, we define for
each muon candidate a variable $T_{iso,\mu}$ \cite{Gray:2011us}
\beq
T_{iso,\mu} = \sum_{0.03 \leq \Delta R_{i,\mu} \leq 0.4}|p_{T, tracks}|_i \times \Theta(|p_{T,tracks}|_i - 0.5 \text{ GeV})
\eeq
where $\Theta (x)$ is the usual Heaviside step function.  Muon
candidates with transverse momentum $p_{T,\mu}$ are identified from an
unmodified PGS simulation of muon chamber hits and are simply required
to satisfy $T_{iso,\mu} / p_{T,\mu} < 0.15$.  Defining $\Delta R_{i,
  \tau}$ to be the distance in $\eta$-$\phi$ space between an object
$i$ and a tau candidate, we define for each tau candidate a variable
$TC_{iso,\tau}$ \cite{Gray:2011us}
\beq
TC_{iso,\tau} = \sum_{0.03 \leq \Delta R_{i,\mu} \leq 0.3} \bigg[ |p_{T,tracks}|_i \times \Theta(|p_{T,tracks}|_i - 0.5 \text{ GeV}) + |E_{T,HCAL}|_i + |E_{T,ECAL}|_i  \bigg]
\eeq
Tau candidates are identified with tracks that have a transverse
momentum $p_{T,\tau}>8$ GeV and must then satisfy $TC_{iso,\tau} /
p_{T,\tau} < 0.15$ for the $\tau$ to be identified.  Tau candidates
which fail this isolation cut are given back to the track list for jet
clustering. Only one-prong hadronic taus are considered for tagging.
Additional correction factors of 0.88, 0.94, and 0.63 are applied to
electron, muon, and tau efficiencies respectively, to better reproduce
the expected CMS lepton counts.  The $b$-tagging routines have been
adapted to approximate the efficiencies and fake rates reported in
\cite{Chatrchyan:2012jua}, and validated on the $b$ and non-$b$ jet
multiplicities as measured in the dileptonic $\ttbar$ cross-section
measurement \cite{CMS-PAS-TOP-12-028}.  The $\ttbar H$ cross-section
is normalized to the NLO results of the LHC Higgs Working Group
\cite{Dittmaier:2011ti}.  The partial widths into $WW^*$ and $\tau^+
\tau^-$ are likewise normalized to the values reported by the LHC
Higgs Working Group (we assume a Higgs mass $m_h = 126$ GeV)
\cite{Dittmaier:2012vm}.

One major advantage of recasting the CMS multi-lepton search is that
most of the backgrounds have already been calculated by the
collaboration, in particular instrumental and heavy-flavour
backgrounds which would be very challenging to otherwise estimate
reliably (though see \cite{Curtin:2013zua} for a proposal).  The
important physics backgrounds $\ttbar W$ and $\ttbar Z^{(*)}$ are
included in the CMS background estimates.  For the $\ttbar h$ signal,
it is also important to consider the further backgrounds $\ttbar W^+ W
^-$ and $\ttbar\ttbar$ \cite{Maltoni:2002jr}, which have not been
included in the experimental studies to date.  We estimate these
backgrounds by generating both processes in MadGraph and both decaying
and showering them in Pythia, before running them through the same
analysis pipeline as the signal.  Cross-sections for $\ttbar W^+ W ^-$
and $\ttbar\ttbar$ are obtained by multiplying the leading order
cross-sections by the same $K$-factor used for $t\bar t H$ production.

Signal and new $\ttbar WW+\ttbar\ttbar$ background predictions are
listed in Table~\ref{tab:20ifb}; the uncertainty on the background
comes from MC error.

\subsection{Limit setting procedure}

To set limits we treat all bins as independent Poisson variables.  To
combine limits from individual bins into a single overall observed
limit we employ a Bayesian algorithm with a flat prior on signal
strength $\mu $, and marginalize over the background uncertainty
according to a log-normal distribution.  We neglect uncertainties on
the theory expectation for Standard Model Higgs cross section and
branching ratios.

To obtain the expected limits, we perform pseudo-experiments in the
following manner: in each signal channel the observed number of events
in the pseudo-experiment is determined by randomly sampling a normal
distribution whose mean and variance correspond to the expected number
of background events and background error. The random number drawn
from this distribution is rounded to the nearest non-negative
integer. This procedure applied to each bin creates one
pseudo-experiment. For each pseudo-experiment, we obtain a limit on
the signal cross section using the same procedure for the observed
limit detailed above.  We perform 400 pseudo-experiments and take the
arithmetic average of the limit from each pseudo-experiment to obtain
the expected limit.

\subsection{Results}

In Table~\ref{tab:20ifb} we show predictions for both $\ttbar H$
signal and the additional backgrounds $\ttbar WW$ and $\ttbar\ttbar $.

\begin{table}[ht]
\centering
\begin{tabular}{|ccc|cccc|}
\hline
  $N_{lep}$ & $N_\tau$ & $S_T$ (GeV) & $t\bar t H$ & $\ttbar WW+\ttbar\ttbar$ & $N_{obs}$ & $N_{pred}$ \\  
  \hline\hline
  3 & 0 & 0-300 & 0.63 & $0.021\pm 0.0045$ & 116 &  $123\pm 50$  \\
  3 & 0 & 300-600 & 4.0 & $0.31 \pm  0.017$ & 130 & $ 127\pm 54$ \\
  3 & 0 & 600-1000 & 1.1 & $0.28 \pm 0.014$ & 13 &  $18.9\pm  6.7$\\
  3 & 0 & 1000-1500 & 0.073 & $0.065 \pm 0.0067$ & 1 & $1.43\pm 0.51$ \\
  3 & 0 & $>$1500 & 0.0027 & $0.0044 \pm 0.0017$  & 0 & $0.208\pm 0.096$ \\
\hline
  3 & 1 & 0-300 & 0.30 & $0.0067\pm 0.0025$ & 710 &  $698\pm 287$ \\
  3 & 1 & 300-600 &  2.1 & $0.12 \pm 0.010$ &  746 & $837\pm 423$\\
  3 & 1 & 600-1000 &  0.57 & $0.12\pm 0.0097$ &  83 &  $97\pm 48$\\
  3 & 1 & 1000-1500 &  0.032 & $0.021\pm 0.0038$ &3 & $6.9\pm  3.9$ \\
  3 & 1 & $>$1500 &  0.0018 & $0.0026\pm 0.0013$ &  0 & $0.73\pm 0.49$\\
\hline
  4 & 0 & 0-300 &  0.029 & $0.00062 \pm 0.00078$ & 0 & $0.186\pm  0.074$ \\
  4 & 0 & 300-600 &  0.16 &  $0.018\pm 0.0038$  &  1 & $0.43\pm  0.22 $\\
  4 & 0 & 600-1000 &  0.036 & $0.013\pm 0.0031$ & 0 & $0.19\pm 0.12$\\
  4 & 0 & 1000-1500 &  0.0019 & $0.0024\pm 0.0013$ &  0 & $0.037\pm 0.0039$ \\
  4 & 0 & $>$1500 &  0.00024 & $0$ & 0 &  $0\pm 0.021$\\
\hline
  4 & 1 & 0-300 &  0.028 & $0.00048\pm 0.00068$ & 1 & $0.89\pm 0.42$\\
  4 & 1 & 300-600 &  0.18 & $0.011\pm 0.0031$  & 0 & $1.31\pm 0.48$\\
  4 & 1 & 600-1000 &  0.052 & $0.014\pm 0.0033$  &  0 & $ 0.39\pm 0.19$ \\
  4 & 1 & 1000-1500 &  0.0043 &$0.0025\pm 0.0013$  & 0 & $  0.019\pm 0.026$\\
  4 & 1 & $>$1500 & $6.6\times 10^{-6}$ & 0  & 0 & $0\pm 0.021$\\
  \hline
  \end{tabular}
  \caption{Number of $\ttbar H$ signal events expected after 19.5
    fb$^{-1}$ at 8 TeV in the multi-lepton bins of
    \cite{Chatrchyan:2013xsw}. Also shown are the number of expected
    $\ttbar WW$ and $\ttbar\ttbar$ events with MC uncertainty,
    together with the number of observed and predicted background
    events reported by CMS.}
  \label{tab:20ifb}
\end{table}

Signal dominantly populates the low-intermediate $S_T$ bin, with $
300\gev\leq S_T< 600$ GeV. The SM backgrounds, especially $\ttbar$,
are also largest in these bins, making $\ttbar H$ a more challenging
signal than higher-mass stops or gluinos. The new backgrounds $\ttbar
WW+\ttbar\ttbar$ are a small fraction of $\ttbar H$ in the lower $S_T$
bins which are most sensitive to $\ttbar H$, but grow rapidly in
importance with $S_T$, numerically dominating over signal in the
highest $S_T$ bins.

The single most sensitive bin is the one with exactly 3 leptons, zero
hadronic taus, and $600\gev\leq S_T< 1000$ GeV.  By itself, this bin
sets a $95\%$ CL upper limit on the signal strength $ \mu \equiv
(\sigma_{tth}\times BR (h \to \mbox{leptons} )) /(
\left.\sigma_{tth}\times BR(h\to\mbox{leptons})) \right|_{SM} < 8.36$.
Combining all bins, we find 
\beq
\mu < 4.7 \: \mathrm{(observed)}, \phantom{scooch}\mu < 6.6 \: \mathrm{(expected)}
\eeq
at $95\%$ CL.  Without the additional backgrounds $\ttbar
WW+\ttbar\ttbar$, observed and expected limits are closer, $\mu_{obs}
< 4.8$ and $\mu_{exp}<6.3$, respectively. Given that the number of
observed events is fixed, the observed limit strengthens with the
inclusion of additional backgrounds as anticipated. 

The two distinct decay modes which populate the multi-leptonic final
state, $h\to WW^*$ and $h\to \tau^+\tau^-$, have nearly identical
contributions in bins with one hadronic tau; in the bins without a
hadronic tau, $h\to WW^*$ contributes at nearly three times the rate
of $h\to\tau^+\tau^-$.  With sufficiently large data sets, such as
those obtained at LHC 14, it would be interesting to separately fit
$WW^*$ and $\tau^+ \tau^-$ profiles in multi-lepton searches.

\section{Same-sign Dileptons}
\label{sec:ssdilep}

Same-sign dileptons are another powerful tool to study rare
leptophilic processes in and beyond the SM.  In this section we
discuss limits on $\ttbar H $ production coming from same-sign
dilepton searches, a complementary strategy to the exclusive three-
and four-lepton searches presented in the previous
section.\footnote{For a recent related proposal to search for $\ttbar
H$ with $H \to WW^*$ at the LHC using same-sign dileptons plus two
$b$-tags, see \cite{Curtin:2013zua}.}

The signal cross-section for same-sign dileptons in $\ttbar H $
production is larger than that in bins with higher lepton
multiplicity, owing to both larger total branching fraction and finite
lepton detection efficiencies. As we discuss below, adding an
exclusive same-sign dilepton bin to the multi-lepton search bins would
significantly enhance the multi-leptonic sensitivity to $\ttbar H$.

To estimate the additional reach provided by a $b$-tagged same-sign
dilepton search for $\ttbar H$, we begin by recasting the CMS search
performed with 10.5 fb$^{-1}$ of 8 TeV data \cite{Chatrchyan:2012paa}.
ATLAS searches in the same final state target signal regions at high
values of $H_T$ ($\met$) which are less populated by SM $\ttbar H$,
though some validation regions offer limited sensitivity
\cite{ATLAS:2013tma, ATLAS-CONF-2013-051}.  Simulation of $\ttbar H$
signal and $\ttbar WW $ and $\ttbar\ttbar $ backgrounds proceeds as in
Sec.~\ref{sec:evtgen}.  Our results are shown in
Table~\ref{tab:ssdilep8}.

\begin{table}
\centering
\begin{tabular}{|l|ccccccc|}
\hline
 & $N_j (N_b)$ & $\met$ (GeV) &  $H_T $(GeV) & $N_{pred} $& $N_{obs} $&  $\ttbar H $& $\ttbar WW +\ttbar\ttbar $  \\
  \hline\hline
SR0 & 2 (2) & $>0$ & $ >80 $ & $40\pm 14$ & 43 & 3.0 &  $0.58\pm 0.015$ \\
SR1 & 2 (2) & $>30$ & $ >80 $ &  $32\pm 11$ & 38 & 2.7 &  $ 0.53 \pm 0.014 $ \\
SR2 & 2 (2) &  $>30$ & $ >80 $ & $17.7\pm  6.1$ & 14 & 1.4 & $  0.27\pm 0.010 $ \\
SR3 & 4 (2) &  $>120$ & $ >200 $ &  $2.2\pm 1.0$ & 1 & 0.023 &  $ 0.045 \pm 0.0033 $\\
SR4 & 4 (2) &  $>50$ & $ >200 $ & $8.1\pm 3.4$ & 10 & 0.073 & $ 0.10 \pm 0.0050 $\\
SR5 & 4 (2) &  $>50$ & $ >320 $ &  $5.7\pm 2.4$ & 7 & 0.068 & $ 0.10 \pm 0.0050 $\\
SR6 & 4 (2) &  $>120$ & $ >320 $ &  $1.7\pm 0.7$ & 1 & 0.023 & $ 0.045 \pm 0.0033 $\\
SR7 & 3 (3) &  $>50$ & $ >200 $ &  $1.2\pm 0.6$ & 1 & 0.21 &$ 0.14 \pm 0.0061 $\\
SR8 & 4 (2) &  $>0$ & $ >320 $ &  $8.1\pm 3.3$ & 9 & 0.089 &$ 0.12 \pm  0.0056$ \\
  \hline
  \end{tabular}
  \caption{Number of $\ttbar H$ signal events expected after 10.5
    fb$^{-1}$ at 8 TeV in the signal regions of
    \cite{Chatrchyan:2012paa}. Also shown are the number of expected
    $\ttbar WW$ and $\ttbar\ttbar$ events with MC uncertainty,
    together with the number of observed and predicted background
    events reported by CMS.  All regions consider both $++$ and $--$
    combinations of leptons, except for SR2, which considers only
    $++$. }
  \label{tab:ssdilep8}
\end{table}

The dominant SM backgrounds to same-sign dilepton searches are $WZ$,
$W^\pm W^\pm q q$, and $\ttbar V$.  Demanding two $b$ tags again
substantially reduces the SM backgrounds.  The remaining backgrounds
are dominated by fake isolated leptons, with physics backgrounds like
$\ttbar V$ secondary \cite{Chatrchyan:2012paa}.  The backgrounds
$\ttbar WW$ and $\ttbar\ttbar$ are again not considered in obtaining
the background estimates, but must be included to assess the reach in
$\ttbar H $.

\subsection{Limits}

In~\cite{Chatrchyan:2012paa} a $ 95\%\: CL_S$ upper bound is quoted on
the number of events from new physics in each of the nine signal
regions for each of three scenarios for the uncertainty on signal rate
times acceptance.  Treating this limit as the bound on $\mu\times
N_{\ttbar H} + N_{\ttbar WW} + N_{\ttbar\ttbar}$ yields upper bounds
on $\mu$, shown in the first row of Table~\ref{tab:ssdilepexcl} (we
show results for an assumed $10\%$ signal uncertainty).  Signal region
SR2, treated as an individual counting experiment, gives a better
single-channel bound than any of the individual bins in the
multi-lepton analysis in the previous section (which use nearly double
the luminosity). For comparison, in the second row of
Table~\ref{tab:ssdilepexcl} we show results for the 95\% CL observed
upper bound in each channel obtained using the limit-setting procedure
in Section~\ref{sec:multilep}.

Unfortunately, the event selection in \cite{Chatrchyan:2012paa} is not
exclusive with respect to to the event selection in the multi-lepton
analysis of \cite{Chatrchyan:2013xsw}, so these limits cannot be
combined with the multi-lepton limits obtained in the previous
section.  Nonetheless, we can get a sense of the additional power an
{\it exclusive} same-sign dilepton bin would add to the leptonic
$\ttbar H$ search by asking what fraction of the signal in the signal
regions of \cite{Chatrchyan:2012paa} arises from events with exactly
two identified leptons according to the multi-lepton identification
criteria of \cite{Chatrchyan:2013xsw}.  This is shown in the third row
of Table~\ref{tab:ssdilepexcl}.  We can then approximate the limit
arising from same-sign dilepton channels {\it exclusive} to the
multi-lepton search procedure by repeating the limit-setting procedure
detailed in Section~\ref{sec:multilep} for the nine same-sign signal
regions, including $t \bar t WW + t \bar t t\bar t$ backgrounds and
``exclusively weighting'' the signal by multiplying the expected
number of signal events in each signal region by the appropriate
exclusive fractions listed in Table~\ref{tab:ssdilepexcl}. The best
limit still arises from SR2, giving an {\it exclusive} limit on $t
\bar t H$
\beq
\mu < 9.2 \: \mathrm{(observed)}, \phantom{scooch}\mu < 13 \: \mathrm{(expected)}
\eeq
at $95\%$ CL. Recalling that the same-sign dilepton search has nearly
half the luminosity of the multi-lepton search, this suggests that the
fraction of same-sign dilepton events exclusive to the multi-lepton
search are still more sensitive to $t \bar t H$ production than the
best single multi-lepton bin.

Finally, we can approximate a combined limit from exclusive same-sign
dilepton and multi-lepton channels using the multi-lepton signal
channels of Section~\ref{sec:multilep} and the best exclusively-weighted
same-sign dilepton signal region (SR2) above. This results in a limit on $t
\bar t H$ from exclusive same-sign dilepton and multi-lepton final
states of
\beq
\mu < 3.8 \: \mathrm{(observed)}, \phantom{scooch}\mu < 5.2 \: \mathrm{(expected)}
\eeq
at $95\%$ CL. This combination should be treated as entirely
approximate due to the non-exclusive nature of the existing
experimental searches in same-sign dilepton and multi-lepton search
channels, which is only roughly compensated for by our exclusive
signal-weighting procedure. Nonetheless, this exercise demonstrates
the potential power of combining a broad range of leptonic final
states with $b$-tags to probe the $t \bar t H$ final state.
Even better sensitivity could be obtained if the same-sign dilepton
exclusive final states were binned by $S_T$, analogously to the
multi-lepton channels.

\begin{table}
\centering
\begin{tabular}{|l|ccccccccc|}
\hline
 & SR0  & SR1 & SR2 & SR3 & SR4 & SR5 & SR6 & SR7 & SR8\\
\hline \hline
$\mu^{95}$ ($CL_S$) & 8.7 & 9.4 & 7.1 & 160 & 150 & 120 & 160 & 17 & 110 \\
$\mu^{95}$ ($CL$) & 11 & 11 & 7.1 & 170 & 140 & 130 & 170 & 19 & 110\\
Excl. Frac. &  0.77 & 0.77 & 0.77 & 0.83 & 0.88 & 0.88 & 0.83 & 0.82 & 0.88 \\
  \hline
  \end{tabular}
  \caption{ First row: modified $CL_S$ limit on $\ttbar H$ production,
    after accounting for new background contributions, as recast from
    \cite{Chatrchyan:2012paa}.  Second row: $CL$ limit on $\ttbar H$
    production, including new background contributions, according to the 
    procedure of Section 2.  Third row:
    fraction of $\ttbar H$ signal events in the signal regions of
    \cite{Chatrchyan:2012paa} which are mutually exclusive with
    respect to the multi-lepton selection criteria of
    \cite{Chatrchyan:2013xsw}.  }
  \label{tab:ssdilepexcl}
\end{table}

We can further use this approximate combination of
exclusively-weighted same-sign dileptons with the multi-lepton search
to estimate the expected sensitivity of the 8 TeV LHC to $\ttbar
H$ production.  Scaling signal and background expectations to 20
fb$^{-1}$, with background uncertainties conservatively taken to scale
linearly with the background, we can approximate the combined reach in
non-resonant leptonic final states to be 
\beq
\mu< 5.0 \; \mathrm{(expected)}
\eeq
This is competitive with measurements in $4b$ and $bb\gamma\gamma$.

\section{Discussion and conclusions}
\label{sec:conclusions}

Direct measurement of the top Yukawa coupling remains one of the key
objectives of the Higgs program at the LHC. In this work we have
demonstrated the viability of measuring $t \bar t H$ associated
production through non-resonant leptonic final states in conjunction
with one or more $b$ tags. These final states arise in $t \bar t H$
production primarily from $H \to \tau^+ \tau^-$ and $H \to WW^*$ in
conjunction with semi-leptonic or fully leptonic decays of the $t \bar
t$ pair.  Recasting the existing CMS $b$-tagged multi-lepton search
with 19.5 fb$^{-1}$ of 8 TeV data, we set an observed (expected) limit
on $t\bar t H$ production of $\mu < 4.7 \; (6.6)$. This sensitivity is
comparable to existing $t \bar t H$ searches targeting $H \to b \bar
b$ and $H \to \gamma \gamma$. 

The multi-lepton limit could be substantially improved by the
inclusion of exclusive channels with same-sign dileptons and two or
more $b$-tags. Although existing search limits in the $b$-tagged
same-sign dilepton final state are not exclusive relative to the
multi-lepton search, we can estimate the gain in sensitivity obtained
with additional exclusive same-sign dilepton channels by recasting the
existing CMS $b$-tagged same-sign dilepton search with 10.5 fb$^{-1}$
of 8 TeV data and approximating the fraction of signal events
exclusive to the multi-lepton selection criteria.  Rescaling same-sign
dilepton signal expectations by this exclusivity factor, we
approximately combine the two searches to set an estimated observed
(expected) limit of 3.8 (5.2).  This limit is already more powerful
than existing measurements in $4b$ and $bb\gamma \gamma$.
Extrapolating the combined expected limits to 20 fb$^{-1}$, we
estimate the sensitivity of the 8 TeV LHC to be $\mu < 5.0$.
Additional sensitivity could be obtained by subdividing the same-sign
dilepton final states into exclusive $S_T$ bins.  Improvement of the
systematic uncertainties on the background predictions with luminosity
would also improve sensitivity relative to our estimate.

We have thus demonstrated that the ability of non-resonant leptonic
searches to observe SM $\ttbar H$ is comparable to the existing search
channels in $4b$ and $bb \gamma \gamma$.  The search proposed here
offers a determination of the top Yukawa coupling which is independent
of the poorly constrained $b$ Yukawa, and of new states in loops, and
is subject to very different experimental systematics.  We expect that
a dedicated search using exclusively defined non-resonant multi-lepton
and same-sign dilepton final states will contribute significantly to
the pursuit of $t \bar t H$ associated production at the LHC.

\acknowledgments Thanks to Y.~Zhao for useful conversations. JS is
supported by NSF grant PHY-1067976 and the LHC Theory Initiative under
grant NSF-PHY-0969510. NC and MP are supported in part by the DOE
under grant DE-FG02-96ER40959. NC is also supported by the NSF under
grant NSF-PHY-0907744 and the Institute for Advanced Study.

\medskip

\noindent {\bf Note added:} While this paper was being completed,
\cite{Onyisi:2013gta} appeared.

\providecommand{\href}[2]{#2}\begingroup\raggedright\endgroup


\end{document}